\documentclass[apj]{emulateapj}
\usepackage{amsmath}
\usepackage{color}
\usepackage{multirow}
\usepackage{threeparttable}

\def\spose#1{\hbox to 0pt{#1\hss}}
\newcommand\lsim{\mathrel{\spose{\lower 3.0pt0pt\hbox{$\mathchar"218$}}
     \raise 2.0pt\hbox{$\mathchar"13C$}}}
\newcommand\gsim{\mathrel{\spose{\lower 3.0pt\hbox{$\mathchar"218$}}
     \raise 2.0pt\hbox{$\mathchar"13E$}}}

\newcommand\mynu{{\nu_{\mathrm{sim}}}}
\newcommand\myupsilon{{\upsilon_{\mathrm{sim}}}}
\newcommand\mysigma{{\sigma_{\mathrm{sim}}}}
\newcommand\mydelta{{\delta_{\mathrm{sim}}}}
\newcommand\myy{{y_{\mathrm{sim}}}}
\newcommand\rth{{R_{\mathrm{th}}}}
\newcommand\gn{{\mathcal{N}_{\mathrm{v2}}}}
\lastpagefootnotes

\usepackage{verbatim}
\usepackage{color}
\usepackage[usenames,dvipsnames]{xcolor}
\definecolor{green}{rgb}{0.0, 0.4, 0.0}
\definecolor{forestgreen(web)}{rgb}{0.13, 0.55, 0.13}
\definecolor{green(web)}{rgb}{0.13, 0.55, 0.13}
\definecolor{green}{rgb}{0.0, 0.4, 0.0}

\def\mnras{Mon.\ Not.\ R.\ Astron.\ Soc.}

\def\aap{Astron.\ Astrophys.}
\def\apj{Astrophys.\ J.}

\def\prd{Phys.\ Rev.\ D}

\def\physrep{Phys. Rep.}

\def\jcap{{JCAP\ }}
\def\jhep{{JHEP\ }}

\begin{document}
%%%%%%%%%%%%%%%%%%%%%%%%%%%%%%%%%%%%%%%%%
\title{New fitting formula for cosmic non-linear density distribution }
\author{Jihye Shin\altaffilmark{1}, Juhan Kim\altaffilmark{2}, Christophe Pichon\altaffilmark{1,3}, Donghui Jeong\altaffilmark{4}, and Changbom Park\altaffilmark{1}}
\altaffiltext{1}{School of Physics, Korea Institute for Advanced Study, 85 Hoegiro, 
Dongdaemun-gu, Seoul 130-722, Korea}
\altaffiltext{2}{Center for Advanced Computation, Korea Institute for Advanced Study, 85 Hoegi-ro, Dongdaemun-gu, Seoul 130-722, Korea}
\altaffiltext{3}{CNRS and UPMC Universit\'{e} Paris 06, UMR 7095, Institut d'Astrophysique de Paris, 98 bis Boulevard Arago, Paris 75014, France}
\altaffiltext{4}{Department of Astronomy and Astrophysics, and Institute for Gravitation and the Cosmos, The Pennsylvania State University, University Park, PA 16802, USA}
%%%%%%%%%%%%%%%%%%%%%%%%%%%%%%%%%%%%%%%%%
\begin{abstract}
 We have measured the probability distribution function (PDF) of cosmic matter 
 density field from a suite of $N$-body simulations. 
We propose the generalized normal distribution of version 2 ($\gn$) as an alternative fitting formula to the well-known log-normal distribution. We find that $\gn$ provides significantly better fit than the log-normal distribution for all smoothing radii ($2$, $5$, $10$, $25$ [Mpc/$h$]) that we studied. The improvement is substantial in the underdense regions. The development of 
 non-Gaissianities in the cosmic matter density 
 field is captured by continuous evolution of the skewness and shifts 
 parameters of the $\gn$ distribution. We present the redshift evolution 
 of these parameters for aforementioned smoothing radii and various
 background cosmology models. All the PDFs measured from large and 
 high-resolution $N$-body simulations that we use in this study can be obtained 
 from a Web site at https://astro.kias.re.kr/jhshin. 

 \end{abstract}
%%%%%%%%%%%%%%%%%%%%%%%%%%%%%%%%%%%%%%%
\section{Introduction}
The inflationary models 
\citep{starobinski:1979,starobinsky:1982,guth:1981,sato:1981,linde:1982,albrecht/steinhardt:1982} 
of the early Universe predict that the 
primordial density perturbations generated during inflation 
\citep{mukhanov/chibisov:1981,hawking:1982,guth/pi:1982,bardeen/etal:1983}
must obey nearly Gaussian statistics
\citep{maldacena:2003,acquaviva/etal:2003,creminelli/zaldarriaga:2004}.
This prediction is confirmed by the observations of temperature anisotropies 
and polarizations of cosmic microwave background radiation 
\citep{planck:NG}, as well as scale-dependent galaxy bias on large-scales 
measured from galaxies \citep{giannantonio/etal:2014} and quasars 
\citep{leistedt/etal:2014}.

The late time nonlinear gravitational evolution, however, induces phase 
coupling in the cosmic matter density and generates non-Gaussian features in 
the one-point probability distribution function (PDF) 
\citep{pee80,jus93,ber94}. The PDFs measured from 
cosmological $N$-body simulations show a significant deviation from the 
Gaussian PDF reflecting the prominent nonlinear 
structures such as clusters, filaments, and cosmic voids \citep{ham85,bou93,kof94,tay00,kay01}.
These late-time non-Gaussian PDFs is directly observable from the cosmic 
shear measurement of weak lensing surveys 
\citep{kruse/schneider:2000,clerkin/etal:2017,takahashi/etal:2011}.
Quantifying the cosmic structure with the non-Gaussian PDF in the cosmic 
density field, therefore, is crucial to understand the nonlinear growth of 
large-scale structure. Upon exploiting the PDFs, one may tighten cosmological 
constraints on, for example, dark energy 
\citep{tatekawa/mizuno:2006,seo/etal:2012,cod16}.

Previous studies on the non-Gaussian PDF have suggested that the distribution 
of the cosmic density field follows approximately the log-normal PDF 
\citep{hub34,col91,kof94,ber95,kay01}. Meanwhile, an alternative fitting formula to the log-normal PDF was proposed by \citet{col94}, the so-called skewed log-normal PDF \citep{ued96}. Since then, pieces of evidence for the deviations from the log-normal distributions have come into sight relying on improved large-box-size, high-precision cosmological simulations \citep{sza04,pandey/etal:2013}. More recently, \citet{uhl16} have analytically calculated 
the deviation of the logarithmic density PDF from the Gaussian one.

In this paper, we propose a new functional form to fit the 
non-Gaussian PDF: the generalized normal distribution of version 2 ($\gn$).
The $\gn$ PDF is a three-parameter extension of the Gaussian distribution 
incorporating the skewness. We show that the PDFs measured from the N-body
simulations are well described by this model over a wide range of density, 
redshift, smoothing kernel radii, and cosmology. 

%%%%%%%%%%%%%%%%%%%%%%%%%%%%%%%%%%%%%%%
\section{Simulation}
We run a suite of cosmological $N$-body simulations using the GOTPM code 
\citep{dub04,kim09,kim11} with $2048^3$ particles in a cubic box of 
$L_{\rm box}=1024~h^{-1}$Mpc. The reference cosmology model 
(hereafter, $\Lambda_{m0}$) adopts the WMAP 5-year cosmology with 
($\Omega_{m,0}$, $\Omega_{b,0}$, $\Omega_{DE,0}$, $w$) = 
($0.26,~0.044,~0.74,-1$), $H_0=72$ km/s/Mpc, and $\sigma_8=0.79$, where $w$ is 
the equation of state parameter of dark energy. Also, we have run four 
simulations with spatially flat, but non-standard cosmologies: 
($\Omega_{m,0}$, $\Omega_{b,0}$, $\Omega_{DE,0}$, $w$) = 
($0.31,~0.044,~0.69,-1$), 
($0.21,~0.044,~0.79,-1$), 
($0.26,~0.044,~0.74,-1.5$), and 
($0.26,~0.044,~0.74,-0.5$) to highlight the effect of total matter density 
and the equation of state of dark energy. We name these simulations, 
$\Lambda_{m-}$, $\Lambda_{m+}$, $Q_{w-}$, and $Q_{w+}$, respectively. 
Each simulation starts from $z=100$ with the initial conditions generated by 
the second-order Lagrangian perturbation theory 
(2LPT; \citealt{scoccimarro:1998,mccullagh/etal:2016,lhu14}) with the linear 
power spectrum calculated from CAMB \citep{lewis/etal:2000}. In this study, we 
have used snapshot particle data at six redshifts of $z=0,~0.2,~0.5,~1,~2$, 
and $4$. 

From this set of particle data, we have measured the one-point PDF on 
$2048^3$ regular grid points laid over the simulation box with the spherical 
top-hat kernel with the radius of $\rth=2,~5,~10$, and $25~h^{-1}$Mpc. The 
particle density is directly measured in real space (direct count in the 
spherical region). 

%%%%%%%%%%%%%%%%%%%%%%%%%%%%%%%%%%%%%%%
\section{One-point density distribution}
%------------------------------------------------------------------------------------------------------ 
 \subsection{Fitting the simulated PDF}
  %-----figure 1-----
\begin{figure}
\centering
\includegraphics{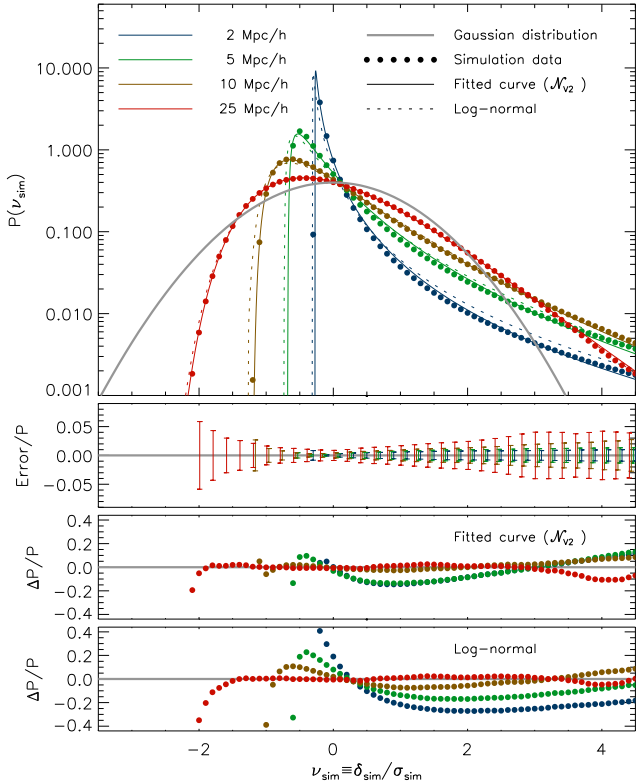}
\caption{$\textit{First panel}$: PDFs with various smoothing radii at $z=0$ in the reference model of $\Lambda_{m0}$. The filled circles are obtained from the simulation ($P_{\mathrm{sim}}$), while the solid lines are the corresponding fitted curves ($P_{\mathrm{fit}}$),  Generalized normal distribution (Ver. 2). The log-normal distribution is over-plotted as the dashed lines. In the legend, we show the radius of the smoothing top-hat kernel. For a reference, we show the Gaussian distribution in the thick gray curve. $\textit{Second panel}$: error bars compared to $P_{\mathrm{sim}}$ to the corresponding smoothing radii in the first panel. For a clear appearance, offsets are used in the x-axis. $\textit{Third and fourth panels}$: relative residuals of $P_{\mathrm{fit}}$ (third) and the log-normal distributions (fourth) compared to $P_{\mathrm{sim}}$ to the corresponding smoothing radii in the first panel. }\label{general}
\end{figure}
%---------------------

After measuring the density, we calculate the probability distribution function
of the density contrast $\delta\equiv\Delta\rho/\langle{\rho}\rangle$ as a function of 
its significance $\mynu$ ($\mynu\equiv\mydelta/\mysigma$ where $\mysigma$ is 
the standard deviation of the density contrast). Hereafter, the subscript of 
`sim' refers to the quantity directly measured from simulation data.  

\begin{figure}
\centering
\includegraphics{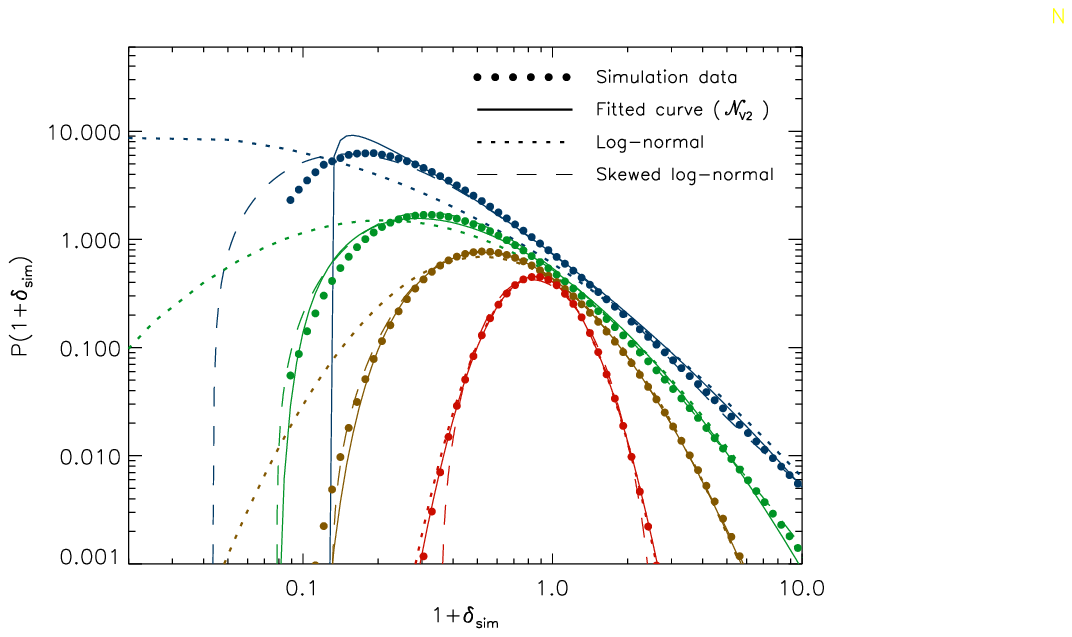}
\caption{Same as top panel of Fig.\ref{general} but as a function of 1+$\mydelta$. The x-axis is scaled in log 1+$\mydelta$ to show the PDFs of the underdense regions in detail. It shows clearly how the fit fails in the underdense nonlinear region (blue curve). }\label{lognormal}
\end{figure}

Figure \ref{general} shows the PDFs measured at $z=0$ from the simulation with
reference cosmology ($\Lambda_{m0}$) for four different 
$\rth=2,~5,~10~,25~{\rm Mpc}/h$. The PDF of density field smoothed with the 
narrower kernels deviates more from the Gaussian distribution (gray, solid 
line); the PDFs are skewed more toward the high-density (right-hand) side and 
present more kurtosis as $\rth$ decreases ($\sigma_{\rm sim}$ gets larger).
To facilitate the comparison in the lower density part of the PDF, we also 
show the same PDFs (but as a function of $1+\mydelta$) in Figure 
\ref{lognormal}. 

We fit the measured PDFs to the generalized normal distribution of version 2 
($\gn$), 
\begin{equation}
\gn(\mynu) = \frac{\mathcal{N}(\myy)}{\alpha+\kappa(\mynu-\xi)},
\end{equation}
in which three parameters $\alpha,~\kappa,~\xi$ are used to parametrize the 
deviation from the normal distribution $\mathcal{N}$.
Here, the distortion argument ($\myy$) is defined as 
\begin{equation}
\myy=\begin{cases}
\displaystyle\frac{1}{\kappa}\ln\left[1+\frac{\kappa(\mynu-\xi)}{\alpha}\right]& \text{if $\kappa \neq 0$}\\
\displaystyle\frac{\mynu-\xi}{\alpha} &\text{otherwise}
\end{cases},
\end{equation}
and $\alpha$, $\kappa$, and $\xi$ quantify, respectively, the scale, shape, 
and location of the skewed distribution $\gn$. A positive (or negative) value 
of $\kappa$ yields left-skewed (or right-skewed) distributions with a sharp 
cut-off in the right (or left) distribution wing. 
Since $\gn$ approaches $\mathcal{N}$ as $\kappa\rightarrow 0$, $\gn$ is useful 
to describe deviations from $\mathcal{N}$ in a continuous manner. In addition,
the cumulative distribution of $\gn$ is the same as that of $\mathcal{N}$ and, 
consequently, $\gn$ is a generalized version of the normal distribution. 
As can be seen in Figure 3, the measured density 
PDFs depend on redshift and smoothing length; therefore, the $\gn$ parameters 
$\alpha$, $\kappa$, and $\xi$ must also be a function of redshift 
(more specifically, the linear growth factor, $D_1$) and $\rth$. 

We find the best-fitting parameters ($\alpha_{\rm fit}$, $\kappa_{\rm fit}$ and
$\xi_{\rm fit}$) by applying the $\chi^2$-minimization method with a thousand density bins.
Hereafter, the subscript of `fit' refers to the best-fitting quantities. 
The resulting best-fitting PDFs to the reference simulation at $z=0$ are 
shown in Figure \ref{general}. As shown there, the overall shape of 
the simulated density PDF ($P_{\mathrm{sim}}$) is well fitted with $\gn$ for a wide range of $\mynu$ and 
$\rth$. 

We also compare $P_{\mathrm{sim}}$ with the log-normal and the skewed log-normal PDFs in Figure \ref{lognormal}. 
The log-normal PDF ($P_{\mathrm{LN}}$)  is defined as
\begin{align}
P_{\mathrm{LN}}(\mydelta)=&\frac{1}{(2\pi\sigma_1^2)^{1/2}}\frac{1}{1+\mydelta}\nonumber\\
	         &\times\exp\left\{\frac{[\ln(1+\mydelta)+\sigma_1^2/2]^2}{2\sigma_1^2}\right\}, 
\end{align}
where the variance $\sigma_1$ can be derived by
\begin{equation}
\sigma_1^2 = \ln[1+\sigma^2_{\mathrm{sim}}].
\end{equation}
The skewed log-normal PDF ($P_{\mathrm{SLN}}$) combines the log-normal distribution and the Edgeworth expansion (e.g. \citealt{jus95}). At third-order approximation \citep{col94}, $P_{\mathrm{SLN}}$ reads
\begin{align}
P_{\mathrm{SLN}}(\myupsilon)  = & \left[1+\frac{1}{3!}T_3\sigma_{\Phi}H_3(\myupsilon)+\frac{1}{4!}T_4\sigma_{\Phi}^2H_4(\myupsilon)\right.\nonumber\\
                                                 & \left.+\frac{10}{6!}T_3^2\sigma_{\Phi}^2H_6(\myupsilon)\right]\mathcal{N(\myupsilon)},
\end{align}
where $\upsilon\equiv\Phi/\sigma_{\Phi}$, $\Phi\equiv\ln(1+\delta)-\langle\ln(1+\delta)\rangle$, and $\sigma_{\Phi}$ is the variance of the log-density field $\Phi$.
$H_m(\upsilon)$ is the Hermite polynomial of degree $m$, and $T_3$ and $T_4$ are the renormalized skewness and kurtosis of the field $\Phi$, respectively:
\begin{equation}
T_3\equiv\frac{\langle\Phi^3\rangle}{\sigma_{\Phi}^4}, T_4\equiv\frac{\langle\Phi^4\rangle-3\sigma_{\Phi}^4}{\sigma_{\Phi}^6}.
\end{equation}
While $P_{\mathrm{sim}}$ is well reproduced 
by all the $\gn$, $P_{\mathrm{LN}}$, and $P_{\mathrm{SLN}}$ in the high-density regions,
the low-density cliffs are better fitted by $\gn$ and $P_{\mathrm{SLN}}$ than by $P_{\mathrm{LN}}$. 
For the smaller $\rth$, the deviation between $P_{\mathrm{sim}}$ 
and $P_{\mathrm{LN}}$ in the underdense regions becomes more 
prominent. It is consistent with the analysis by 
\citet{ued96} and the perturbation theory presented by \citet{ber95}. Although
the fits by $\gn$ also differ from $P_{\mathrm{sim}}$ in the underdense regions,
in particular for the smaller $\rth$, the deviation is much milder than that 
of $P_{\mathrm{LN}}$.

  %-----figure 2-----
%---------------------
 %-----figure 3-----
\begin{figure*}
\includegraphics{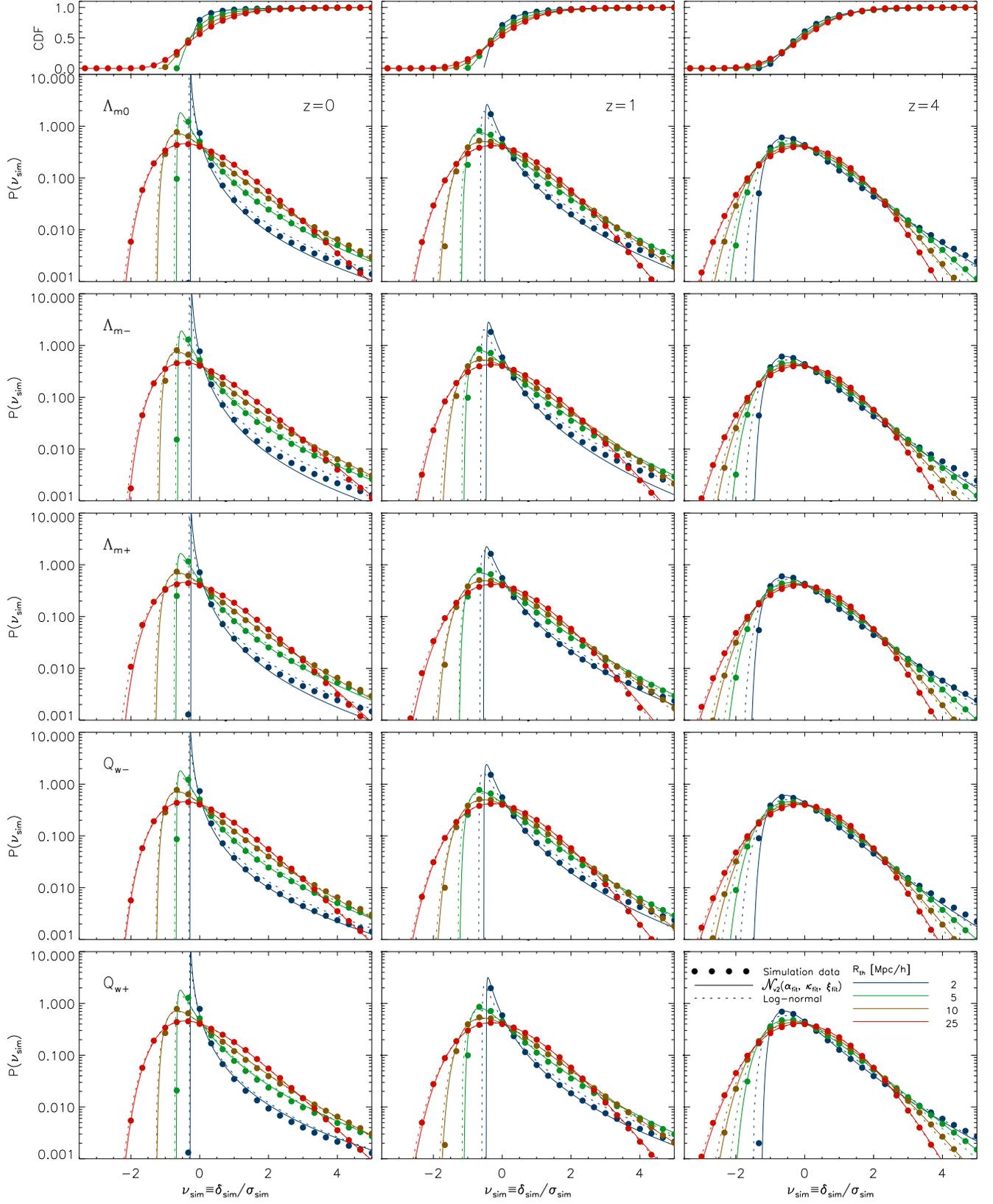}
\caption{PDFs with various smoothing radii at $z=$ 0, 1, and 4 for the five simulated models. $P_{\mathrm{sim}}$ (filled circles) are compared with the corresponding predicted PDFs,  $\gn(\alpha_{\mathrm{fit}}, \kappa_{\mathrm{fit}}, \xi_{\mathrm{fit}})$ (solid lines), and the log-normal distributions (dashed lines). In the first row, cumulative distribution functions (CDFs) of $P_{\mathrm{sim}}$ (filled circles) are compared with the corresponding CDFs of  $\gn(\alpha_{\mathrm{fit}}, \kappa_{\mathrm{fit}}, \xi_{\mathrm{fit}})$, which are the same to the CDFs of $\mathcal{N}(\myy)$. }
\label{all}
\end{figure*}
%---------------------
  %-----figure 4-----
\begin{figure}
\centering
\includegraphics{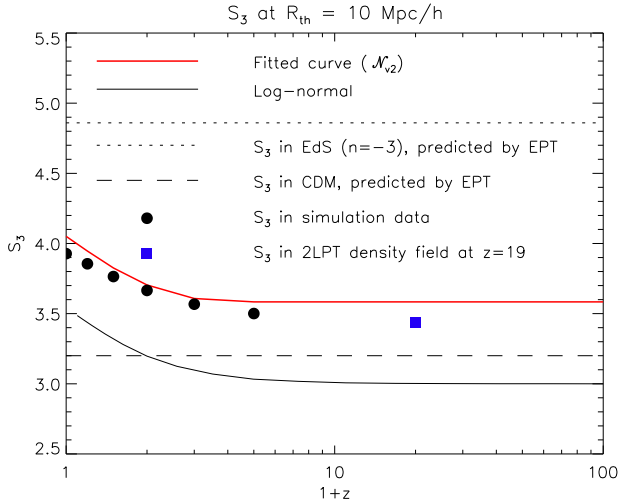}
\caption{$S_3$ values of $\gn(\alpha_{\mathrm{fit}}, \kappa_{\mathrm{fit}}, \xi_{\mathrm{fit}})$ (red line) and $P_{\mathrm{LN}}$ (black line) as a function of 1+z. Horizontal lines indicate $S_3$ values in EdS universe with $n=-3$ (dotted line) and in the CDM universe (dashed line), predicted by the EPT. Black filled circles represent $S_3$ values measured from our simulation data, while blue filled square indicates that from the 2LPT density field at $z=19$. Here, the same smoothing radius as $\rth=10$~Mpc/h is adopted for all the $S_3$ values.}\label{skewness}
\end{figure}
%---------------------
 %-----figure 5-----
\begin{figure*}
\includegraphics{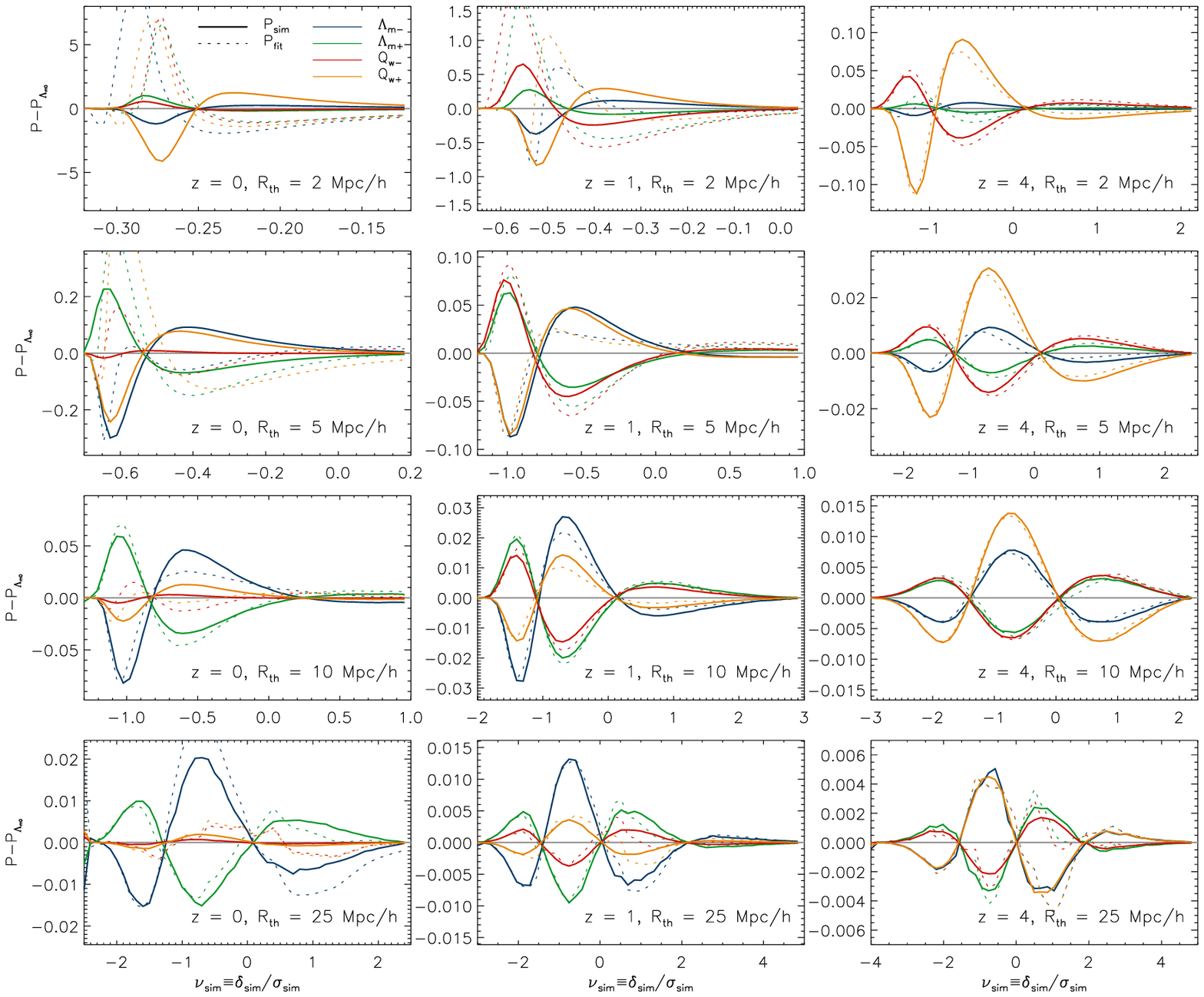}
\caption{Difference of PDFs ($P-P_{\Lambda_{m0}}$) for various simulated models ($\Lambda_{m-}$, $\Lambda_{m+}$, $Q_{w-}$, $Q_{w+}$) relative to the reference model ($\Lambda_{m0}$). The thick solid lines are obtained from the simulations ($P_{\mathrm{sim}}$), while the dashed lines are from the corresponding fitted curves by $\gn$ ($P_{\mathrm{fit}}$). For the smaller radii and the lower redshift, the model fails to capture the difference between the different cosmologies and the reference $\Lambda$CDM model.}
\label{cosmologies}
\end{figure*}
%---------------------
%----------------------------------------------------------------------------------------------------- 
\subsection{Fitting to $\alpha$, $\kappa$, and $\xi$}

All best-fitting parameters ($\alpha_{\rm fit}$, $\kappa_{\rm fit}$, and 
$\xi_{\rm fit}$) vary with redshifts (or $D_1$) and $\rth$. We therefore 
compile these fitting values at six redshifts ($z=$~0, 0.2, 0.5, 1, 2, and 4), 
four smoothing radii ($\rth=$~2, 5, 10, and 25~$h^{-1}$Mpc) for five different 
simulated models ($\Lambda_{m0}$, $\Lambda_{m-}$, $\Lambda_{m+}$, $Q_{w-}$, 
and $Q_{w+}$). We then find functions that incorporate the redshift 
and smoothing-scale dependence of the best-fitting parameters by using the 
Eureqa software \footnote{http://www.nutonian.com/products/eureqa}. 

Among all possible fitting forms to $\alpha_{\mathrm{fit}}(\rth,D_1)$, $\kappa_{\mathrm{fit}}(\rth,D_1)$, and $\xi_{\mathrm{fit}}(\rth,D_1)$, we select those which satisfy the following criteria: 
(1) the functional form must be the same for all the cosmological models, 
(2) the resulting PDF must asymptote to the normal distribution 
at early times and for large smoothing radius;
that is, $\alpha_{\mathrm{fit}}\to 1$, $\kappa_{\mathrm{fit}}\to 0$ and 
$\xi_{\mathrm{fit}}\to 0$ as $\rth\rightarrow\infty$ and/or $D_1\rightarrow0$, (3) the R$^2$ goodness of fit should be larger than $0.9$, and (4) the fitting equation has the least number of coefficients. 
Note that the empirical relations that we find here do not necessarily reflect 
the physical origin of the underlying function.
The final functions that we find are
\begin{eqnarray}
\alpha_{\rm fit}(\rth, D_1)  & = & \frac{\rth^2}{\rth^2 +a_1 D_1^2+\rth D_1^2},\\ 
\kappa_{\rm fit}(\rth, D_1) & = & \frac{b_1 D_1}{b_2+\rth+b_3 D_1},\\ 
\xi_{\rm fit}(\rth, D_1)        & = & \frac{c_1 \rth D_1}{\rth+\rth^2+c_2 D_1^2},\\
\end{eqnarray}
respectively, where $a_i$, $b_i$, and $c_i$ are numerical coefficients.

Table \ref{tbl-1} lists the coefficients and their $R^2$ goodness of fit value for the simulated models. The predicted $\gn$ by $\alpha_{\mathrm{fit}}(\rth,D_1)$, $\kappa_{\mathrm{fit}}(\rth,D_1)$, and $\xi_{\mathrm{fit}}(\rth,D_1)$, hereafter $\gn(\alpha_{\mathrm{fit}}, \kappa_{\mathrm{fit}}, \xi_{\mathrm{fit}})$,  are compared with the corresponding $P_{\mathrm{sim}}$ and the log-normal distribution in Figure \ref{all}. $\gn(\alpha_{\mathrm{fit}}, \kappa_{\mathrm{fit}}, \xi_{\mathrm{fit}})$ reproduce well the overall shape of $P_{\mathrm{sim}}$  for a wide range of $\mynu$, $\rth$, redshift, and cosmology. The PDFs over the entire density scale are better reproduced by $\gn(\alpha_{\mathrm{fit}}, \kappa_{\mathrm{fit}}, \xi_{\mathrm{fit}})$ than with the log-normal distribution. 

%-----table 1-----
\begin{table}[ht!]
\caption{Numerical coefficients of Fitted functions for $\alpha$, $\kappa$, and $\xi$ \label{tbl-1}}
\begin{tabular}{clc}
\tableline\tableline
Model & Numerical coefficients  &$R^2$ goodness\\
\tableline
\tableline
\multirow{3}{*}{$\Lambda_{m0}$} 
& $a_1=32.92 $&0.992 \\
& $b_1=10.38, b_2=2.118, b_3=3.151$ & 0.994\\
& $c_1=-4.082, c_2=44.92$ & 0.969 \\
\tableline
\multirow{3}{*}{$\Lambda_{m-}$} 
& $a_1=35.48$&0.992 \\
& $b_1=11.85, b_2=3.132, b_3=3.683$ & 0.993\\
& $c_1=-4.370, c_2=54.48$ & 0.985 \\
\tableline
\multirow{3}{*}{$\Lambda_{m+}$} 
&$a_1=30.82$ & 0.992\\
&$b_1=9.369, b_2=1.426, b_3=2.787$ & 0.993\\
&$c_1=-3.901, c_2=39.85$ & 0.972\\
\tableline
\multirow{3}{*}{$Q_{w-}$} 
&$a_1=31.54$ & 0.992 \\
&$b_1=10.22, b_2=2.920, b_3=2.124$& 0.996 \\
&$c_1=-3.929, c_2=41.41$ & 0.980\\
\tableline
\multirow{3}{*}{$Q_{w+}$} 
&$a_1=32.05$ & 0.994\\
&$b_1=10.30, b_2=2.890, b_3=2.250$ &0.996 \\
&$c_1=-4.024, c_2=45.35$ &0.980\\
\tableline
\tableline
\end{tabular}
\end{table}
%---------------------
%------------------------------------------------------------------------------------------------------ 
\subsection{Skewness of fitted PDFs}
The density fluctuations in the very early universe are known to be indistinguishable from Gaussian to within measurement error. However, gravity is expected to skew the density distribution, making a lognormal, skewed lognormal, or $\gn$ a better fit than a Gaussian even at early times \citep{pee80,fry84,jus93,ber94}. It should be noted, though, that for small variance at early times, the skewness has a negligible effect on the actual density PDF. Eulerian perturbation theory (EPT) predicts that a reduced skewness parameter $S_3\equiv\left<\delta^3\right>/{\sigma}^4$  in an Einstein de sitter (EdS) universe approaches to $\sim34/7-(n+3)$ at early times, where $n$ is an index of power-law spectrum \citep{pee80,jus93,ber94,fry94}. Since the log-normal PDF has a non-zero skewness as $S_3=3$ at early times, the log-normal distribution has been proposed as a better fit than a Gaussian to the initial PDF \citep{col91,col94,ney13}. Figure \ref{skewness} compares the $S_3$ values of $\gn(\alpha_{\mathrm{fit}}, \kappa_{\mathrm{fit}}, \xi_{\mathrm{fit}})$ and $P_{\mathrm{LN}}$ for the $\Lambda$CDM model ($\Lambda_{m0}$) at $R_{\mathrm{th}} = 10$ Mpc/h.  Although the $\gn(\alpha_{\mathrm{fit}}, \kappa_{\mathrm{fit}}, \xi_{\mathrm{fit}})$ is chosen to converge to the normal distribution at early times (second condition of \S 3.2), the $S_3$ value approach to a non-zero value of $\sim3.6$. The $S_3$ values directly measured from our simulation data (black circles) closely follow that of $\gn(\alpha_{\mathrm{fit}}, \kappa_{\mathrm{fit}}, \xi_{\mathrm{fit}}$) rather than of $P_{\mathrm{LN}}$. To calculate the $S_3$ value at the higher redshift, we generate an initial Gaussian density field at $z=19$ evolved by the 2LPT. Following the $S_3$ trend of the simulation data, the $S_3$ value at $z=19$ (blue filled square) results in $S_3\sim3.4$, which is slightly smaller than that of $\gn(\alpha_{\mathrm{fit}}, \kappa_{\mathrm{fit}}, \xi_{\mathrm{fit}}$).
%------------------------------------------------------------------------------------------------------ 

\subsection{Sensitivity of fitted PDFs to cosmology}

Relative differences of PDFs for four non-standard models ($\Lambda_{m-}$, $\Lambda_{m+}$, $Q_{w-}$, $Q_{w+}$) relative to the $\Lambda$CDM model ($\Lambda_{m0}$) are shown in Figure \ref{cosmologies}. The differences ($P-P_{\Lambda_{m0}}$) compiled by both $P_{\mathrm{sim}}$ and $P_{\mathrm{fit}}$ are compared to each other in order to check how $P_{\mathrm{fit}}$ capture the different models.  $P-P_{\Lambda_{m0}}$ are well reproduced at high redshift and/or the large smoothing. However, $P-P_{\Lambda_{m0}}$ for smaller redshift or the small smoothing show significant deviations from that of $P_{\mathrm{sim}}$. Thus, the $\gn$ fits are not accurate enough to make the distinction between the models for the strongly nonlinear regime. The failure is due to poor fits of the PDFs in the underdense region (see Fig. \ref{lognormal}).

%%%%%%%%%%%%%%%%%%%%%%%%%%%%%%%%%%%%%%
\section{Summary \& Discussion}

In this paper, we presented the one-point PDFs measured from cosmological 
$N$-body simulations and showed that the new fitting formula based on the 
generalized normal distribution of version 2 ($\gn$) provides significantly 
better fit compared to the log-normal distribution. 
In particular, $\gn$ reproduces well the overall PDFs for a wide range of 
density, smoothing kernel, redshift, and cosmology, except in strongly 
nonlinear regimes. 
The improvement by the $\gn$ is substantial in the under-dense regions, which is also achieved by the skewed log-normal distribution, the third-order Edgeworth expansion of the log-normal distribution \citep{col94,ued96}.

As the $\gn$ distribution can accommodate a continuous transition from the 
initial Gaussian distribution function, the result we present here should 
pave the  way to modeling the density PDF in the quasi-linear regimes
where perturbation theory \citep{bernardeau/etal:2002} captures the 
nonlinear evolution of cosmic density fields.

The simulated PDFs and their fitted curves by $\gn$ for 
various smoothing kernels ($\rth=2$, 5, 10, and 25 $h^{-1}$Mpc), redshifts 
($z=0$, 0.2, 0.5, 1, 2 and 4), and cosmologies ($\Lambda_{m0}$, $\Lambda_{m-}$, 
$\Lambda_{m+}$, $Q_{w-}$, and $Q_{w+}$) can be obtained from the Web site of 
the first author at https://astro.kias.re.kr/jhshin. 

%%%%%%%%%%%%%%%%%%%%%%%%%%%%%%%%%%%%%%%
\acknowledgements
We appreciate the anonymous referee for his/her helpful comments that improved our manuscript. We thank the Korea Institute for Advanced Study for providing computing resources (KIAS Center for Advanced Computation) for this work. We thank C. Uhlemann for comments. CP's research is partially funded by  Spin(e) ANR-13-BS05-0005. DJ acknowledges support from National Science Foundation grant AST-1517363.

%%%%%%%%%%%%%%%%%%%%%%%%%%%%%%%%%%%%%%%

%%%%%%%%%%%%%%%%%%%%%%%%%%%%%%%%%%%%%%%
\end{document}